\documentclass[11pt,a4paper,iop,apj]{emulateapj}
\usepackage{epsfig}
\shorttitle{The distance to the Small Magellanic Cloud}
\shortauthors{Graczyk, Pietrzy{\'n}ski et al.}
\begin{document}
\title{The Araucaria Project. An accurate distance to the late-type double-lined eclipsing binary OGLE SMC113.3 4007 in the Small Magellanic Cloud\altaffilmark{$\star$}}
\author{Dariusz Graczyk\altaffilmark{1}, Grzegorz Pietrzy{\'n}ski\altaffilmark{1,2}, Ian B. Thompson\altaffilmark{3}, 
Wolfgang Gieren\altaffilmark{1}, Bogumi{\l} Pilecki\altaffilmark{1,2}, Andrzej Udalski\altaffilmark{2}, 
Igor Soszy{\'n}ski\altaffilmark{2}, Zbigniew Ko{\l}aczkowski\altaffilmark{4}, Rolf-Peter Kudritzki\altaffilmark{5,10}, Fabio Bresolin\altaffilmark{5}, Piotr Konorski\altaffilmark{2}, Ronald Mennickent\altaffilmark{1}, 
Dante Minniti\altaffilmark{6}, Jesper Storm\altaffilmark{7}, Nicolas Nardetto\altaffilmark{8} and Paulina Karczmarek\altaffilmark{9}}
\altaffiltext{$\star$}{Based on observations obtained with the ESO NTT and 3.6m telescopes at La Silla, and with the Magellan Clay telescope at Las Campanas Observatory}
\affil{$^1$Universidad de Concepci{\'o}n, Departamento de Astronom{\'i}a, Casilla 160-C, Concepci{\'o}n, Chile; darek,wgieren,bpilecki,rmennick@astro-udec.cl}
\affil{$^2$Obserwatorium Astronomiczne, Uniwersytet Warszawski, Al.~Ujazdowskie 4, 00-478, Warszawa, Poland; pietrzyn,udalski,soszynsk,piokon@astrouw.edu.pl}
\affil{$^3$Carnegie Observatories, 813 Santa Barbara Street, Pasadena, CA 911101-1292, USA; ian@obs.carnegiescience.edu}
\affil{$^4$Instytut Astronomiczny, Uniwersytet Wroc{\l}awski, Kopernika 11, 51-622, Wroc{\l}aw, Poland; kolaczkowski@astro.uni.wroc.pl}
\affil{$^5$Institute for Astronomy, University of Hawaii at Manoa, 2680 Woodlawn Drive, Honolulu, HI 96822, USA; kud,bresolin@ifa.hawaii.edu}
\affil{$^6$Departamento de Astronomia y Astrofisica, Pontificia Universidad Catolica de Chile, Casilla 306, Santiago 22, Chile; dante@astro.puc.cl}
\affil{$^7$Leibniz Institute for Astrophysics, An der Sternwarte 16, 14482 Potsdam, Germany; jstorm@aip.de}
\affil{$^8$Laboratoire Lagrange, UMR7293, UNS/CNRS/OCA, 06300 Nice, France; Nicolas.Nardetto@oca.eu}
\affil{$^9$Centrum Astronomii, Uniwersytet Miko{\l}aja Kopernika, Gagarina 11, 87-100 Toru{\'n}, Poland; paulina.karczmarek@gmail.com}
\affil{$^{10}$Max-Planck-Institute for Astrophysics, 
Karl-Schwarzschild-Str.1, D-85741 Garching, Germany} 

\begin{abstract}
We have analyzed the long period, double-lined eclipsing binary system OGLE SMC113.3 4007 (SC10 137844) in the SMC. The binary lies in the north-eastern part of the galaxy and consists of two evolved, well detached, non-active G8 giants. 
The orbit is eccentric with $e=0.311$ and the orbital period is  371.6 days. Using extensive high-resolution spectroscopic and multi-color 
photometric data we have determined a true distance modulus of the system of m-M=18.83 $\pm$ 0.02 (statistical) $\pm$ 0.05 (systematic) mag
using a surface brightness - 
color relation for giant stars. 
This method is very insensitive to metallicity and reddening corrections and depends only very little on stellar atmosphere model
assumptions. Additionally, we derived very accurate, at the level of 1\%-2\%, physical parameters of both giant stars,
particularly their masses and radii, making our results important for comparison with stellar evolution models. Our analysis underlines the high potential of late-type, double-lined detached binary systems for accurate distance determinations to nearby galaxies.   
\end{abstract} 

\keywords{binaries: eclipsing --- galaxies: distances and redshifts --- galaxies: individual (\objectname{SMC}) --- stars: late-type} 
\section{Introduction}
The determination of the cosmic distance scale is one of the fundamental tasks in astronomy. Perhaps
it is the most fundamental one because usually physical  quantities used in astrophysics 
are scaled by the distance. 
The building of the cosmic distance ladder has proven to be a difficult, time consuming multi-step task. The first step
consists of using geometrical methods to measure the distances of nearby stars and clusters.
There are only a few well defined, purely geometrical methods we can use: 1) the annual trigonometric parallax, 
2) the proper motions convergence point of a star cluster, 3) the orbital parallax of a spectroscopic visual binary star, 
4) the kinematic parallax of a pulsar. These primary methods are currently limited to
 only a few hundred parsecs around the sun. In some favorable 
circumstances this limit can be expanded to a few kiloparsecs: examples are the distance determination to the Galactic Center via 
the orbital parallax of stars using adaptive optics e.g.,~\citet{gil09} and the annual trigonometric parallax determination 
of Milky Way masers using long-baseline radio interferometry e.g.,~\citet{mol09,red09,nag11}. To reach out beyond
the Milky Way we have to use secondary methods calibrated with the help of the primary ones, 
or use other direct but "near-geometrical" methods based on some astrophysical models. In the first case we are restricted quite severely by the
limited volume of space where we can safely calibrate secondary distance indicators. In the latter case we are affected 
by the model assumptions inherent to the different methods. Additionally, these model assumptions are often based on an
accepted distance scale and may produce a circular reasoning - see as an example the discussion of the Pleiades' 
distance modulus problem by \citet{lee09}.

A fundamental step in the construction of the distance ladder is the determination of accurate distances to the Magellanic Clouds, particularly
to the Large Magellanic Cloud. This galaxy with its relatively simple geometrical structure and smaller than Small Magellanic Cloud depth in the line of sight is a rich astrophysical laboratory containing sizable samples of stellar distance indicators of different kinds and still remains an ideal anchor point for the distance scale (see recent review of \cite{wak11}).
One of the fundamental tasks
of the Araucaria project is to optimize existing methods by checking on their systematic errors.
We have been using red clump stars \citep{pie02,pie03}, Cepheid variables \citep{gie05,sto11a,sto11b} 
and RR Lyrae stars \citep{sze08}
to measure the LMC distance, with all these techniques taken to the near-infrared to minimize systematic errors due
to dust extinction on the results. These different methods have yielded true LMC distance moduli in the range 18.45 - 18.58 mag, with total
errors in the range 4-6\%. While this work has already yielded important progress in the long-standing effort of determining
a truly accurate distance to the LMC, it is desirable and necessary for precision cosmology, and in particular for the determination of
an accurate (better than 5\%) value of the Hubble constant, to do better and determine the LMC distance to 1\%. A very
encouraging step in this direction was recently made in our project with the analysis of a first late-type eclipsing binary in the LMC,
OGLE-051019.64-685812.3. This detached, double-lined binary system consisting of two G4 giant stars has allowed a near-geometrical distance measurement accurate to 3\% \citep{pie09}. We are currently observing and analyzing 
a number of similar systems in the LMC from which we expect to measure the LMC barycenter distance with total uncertainty better than 2.5\%. We have also detected
several detached, double-lined eclipsing binary systems with red giant components in the SMC which are extremely useful for a precision distance measurement to that galaxy. 

There are a large number of implementations of the eclipsing binary (EB) method in the literature. 
They differ mostly in amount of the modeling needed to calculate the total emergent flux of a binary system. 
The computed flux is scaled to the observed flux at Earth to get a distance. One implementation 
employs stellar atmosphere models to calculate the emergent flux by fitting them to the observed spectral energy 
distribution (SED) obtained with the spectrophotometry. Such an approach was undertaken by E.~Guinan and collaborators 
in the middle of 90's \citep{gui96} with the goal to obtain a precise distance to the LMC with a total uncertainty 
of about 2\%-3\%. They studied EBs with late O-type and early B-type components bright enough to obtain 
sufficient quality data \citep{gui98a}. After, their pioneering work on the distance determination to the early type 
LMC binary HV 2274, with a claimed accuracy of 3.5\% \citep{gui98b}, it had become clear that the method has indeed a 
large potential. However, the extensive use of stellar atmosphere models in the fitting procedure makes such an approach 
very model dependent. Because absolute fluxes, especially in the UV region, predicted by atmosphere models for very 
early type stars can be calibrated with an accuracy of at most 5\% (principally caused by the lack of a sufficient number 
of nearby calibrator stars with reliable geometrical parallaxes), the interpretation of the results obtained from early type 
systems is uncertain, e.g. \citet[see their section 7]{fit03} in the case of the LMC and \citet[see their section 5.6]{nor10} 
in the case of the SMC.        

A different possibility is to employ an empirical surface brightness - color relation and to use angular diameter scaling 
to obtain a distance to an eclipsing binary. Such relations, for different colors, are well established for dwarfs 
and giants with spectral types later than F0 from accurate measurements of stellar angular diameters by 
interferometry \citep{gro04,ben05,ker08,cas10}. This approach was used by us to determine the distance to  
OGLE-051019.64-685812.3 in the LMC \citep{pie09}. Although the 1\% precision of distance determination 
to extragalactic eclipsing binary systems envisioned by \citet{bog97} is, in principle, possible to obtain, 
we are presently still hampered by the systematic uncertainty on the surface brightness-color calibration - see the section 4.1 of this paper.  

In this paper we present the distance determination to the long period, well detached eclipsing binary OGLE SMC113.3 4007 ($\alpha_{2000}=01\!^{\rm h}06^{\rm m}$10\fs33; $\delta_{2000}=-$72\arcdeg06\arcmin25\farcs2) in the SMC. As this system is ideal for distance determination the distance, we show the full potential of the method and discuss all possible sources of error which can lower the robustness of our final result. The star was identified  as an eclipsing binary in the catalogue published by \cite{uda98} and obtained identifier SC10 137844 in the OGLE-II database. Initially, because the eccentricity of the system is close 
to one third and the secondary minimum appears almost exactly at orbital phase 0.67, the system was incorrectly assumed to have a circular orbit and a period of $\sim$ 248 days. \cite{wyi01} selected SC10 137844 as a prime  distance indicator to the SMC and shortly afterwards we started obtaining high-resolution spectroscopic observations of the star. The system was identified as an eclipsing binary in the MACHO database by \cite{fac07} as MACHO 206.17005.6. In the OGLE-III photometric database the star is listed as SMC113.3 4007.

\section{Observations and Data Reduction}
Johnson-Cousins optical photometry of the system was obtained with the Warsaw 1.3~m telescope at Las Campanas Observatory in the course 
of the second, third and fourth phases of the OGLE project \citep{uda97,uda03} and also with the ANDICAM camera attached to the CTIO 1.3~m telescope. 
In total we secured 1178 I-band epochs and 168 V-band epochs. Because of the long orbital period consecutive epochs were taken usually 
on different nights. The time span of the I-band observations is 5119 days (from JD  2450627 to JD 2455746). The raw data were reduced with the image-subtraction 
technique \citep{woz00,uda03} and instrumental magnitudes were calibrated onto the standard system using Landolt standards. We did not use MACHO data for this star because OGLE I-band photometry has better eclipse coverage and better precision (smaller noise and much smaller number of outliers).  
\begin{figure}
\includegraphics[angle=0,scale=.50]{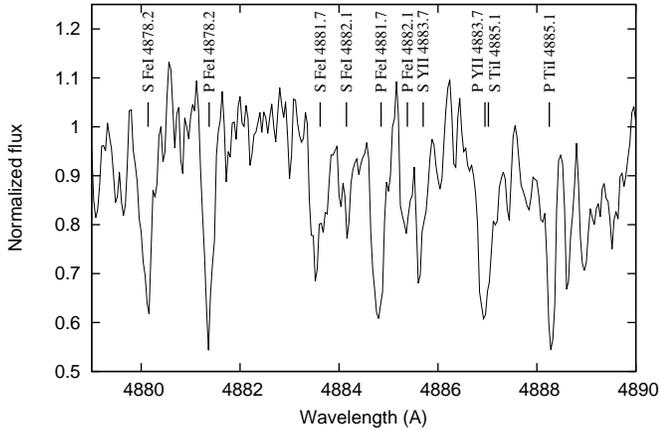}
\caption{ A MIKE spectrum of OGLE SC10 137844 taken at the third quadrature. Some absorption lines from both components are 
identified. \label{fig0}}
\end{figure}

Near-infrared photometry was collected with the ESO NTT telescope on La Silla equipped with the SOFI camera. 
The setup of the instrument, and the reduction and calibration of the data onto the UKIRT system were essentially identical to those 
described in our previous paper by \cite{pie09}. The transformation of our photometry onto the Johnson system was done using the equations given 
by \cite{car01} and \cite{bes88}. 

High resolution echelle spectra were collected with the Clay 6.5~m telescope at Las Campanas Observatory, equipped with the MIKE spectrograph, 
and with the 3.6~m telescope on ESO La Silla, equipped with the HARPS spectrograph. We employed an $5\times0.7$ arc sec slit with MIKE giving 
a spectral resolution of about 40000.  In the case of HARPS we used the EGGS mode yielding a resolution of about 80000.  In our analysis 
we used 8 HARPS spectra, 16 MIKE blue region and 14 MIKE red region spectra (25 epochs; 13 times blue and red MIKE spectra were taken simultaneously). The typical S/N at $\sim$5000~\AA$\,$ was about 20 and 10 for MIKE 
and HARPS spectra, respectively. Figure~\ref{fig0} shows a small part of a MIKE blue region spectrum taken close to the third quadrature 
(JD 2454816.6).

Before further analysis all the spectra were shifted to the Solar System Barycenter. In order to derive radial velocities of the system's components
 we employed the Broadening Function (BF) formalism \citep{ruc92,ruc99} and the two-dimensional cross-correlation method TODCOR 
 by \cite{maz94}. We used numerous metallic lines in the wavelength regions 4125-4230, 4245-4320, 4350-4840, 4880-5000, 5350-5850, 5920-6250, 
 6300-6390, 6600-6800~\AA. As templates we used synthetic spectra from a library computed by \cite{col05}. The templates were chosen 
 to closely match the atmospheric properties of the stars in a grid of $T_{\rm eff}$ and $\log g$. We iterated the radial velocity determination 
 a few times each time updating the template according to the latest solution. Usually the difference between velocities derived by TODCOR and BF 
 was smaller than 100 m~s$^{-1}$ and we adopted the mean value of both as final measurement.  We detected a linear trend in velocities derived 
 from MIKE spectra taken between HJD 2454000 and 2454800 equal to about $-0.6$ m~s$^{-1}$ per day.  After removing the trend we accounted also, 
 during the mentioned period of time, for a zero point shift between the MIKE and HARPS spectrographs of 0.344 km~s$^{-1}$. The individual 
 RV measurements are listed in Table~\ref{tbl-1}.

\begin{deluxetable}{lcclcc}
\tabletypesize{\scriptsize}
\tablecaption{Radial Velocity Measurements for SC10 137844 \label{tbl-1}}
\tablehead{
\colhead{HJD} & \colhead{$V_1$} & \colhead{$V_2$} & \colhead{HJD} & \colhead{$V_1$} & \colhead{$V_2$}  \\
\colhead{-2450000} & \colhead{km s$^{-1}$} & \colhead{km s$^{-1}$} & \colhead{-2450000} & \colhead{km s$^{-1}$} & \colhead{km s$^{-1}$} 
}
\startdata
 2946.65112\tablenotemark{$\dagger$}  &  183.85  &  110.99 &  4648.92647\tablenotemark{$\dagger$}   &  129.83  &  164.94 \\
 4004.71811  &  157.44  &  134.47 &  4648.92647  &  130.16  &  165.14 \\
 4004.71812\tablenotemark{$\dagger$}   &  157.44  &  134.82 &  4655.93369  &  131.01  &  164.06 \\
 4010.72315  &  159.90  &  133.69 &  4655.93375\tablenotemark{$\dagger$}   &  131.15  &  164.08 \\
 4010.72315\tablenotemark{$\dagger$}   &  160.50  &  133.67 &  4656.87817\tablenotemark{$\dagger$}   &  130.92  &  163.75 \\
 4065.70091  &  185.64  &  111.32 &  4656.87817  &  131.00  &  163.61 \\
 4065.70093\tablenotemark{$\dagger$}   &  184.75  &  111.06 &  4671.91268\tablenotemark{$\star$}  &  133.38  &  160.74 \\
 4314.88758\tablenotemark{$\dagger$}   &  137.63  &  157.86 &  4686.79155\tablenotemark{$\star$}  &  137.42  &  157.33 \\
 4314.89337  &  137.13  &  157.35 &  4689.91736  &  138.86  &  157.30 \\
 4316.88748  &  138.02  &  157.37 &  4698.85661\tablenotemark{$\star$}  &  140.75  &  154.99 \\
 4329.86296\tablenotemark{$\dagger$}   &  141.33  &  154.28 &  4808.54742\tablenotemark{$\star$}  &  185.18  &  111.22 \\
 4329.86296  &  141.90  &  154.61 &  4809.54221\tablenotemark{$\star$}  &  185.02  &  110.64 \\
 4395.64368\tablenotemark{$\star$}  &  166.57  &  129.13 &  4816.56866  &  185.83  &  109.75 \\
 4412.54544  &  175.60  &  120.99 &  4816.56866\tablenotemark{$\dagger$}   &  185.85  &  110.18 \\
 4419.59240\tablenotemark{$\star$}  &  178.64  &  117.33 &  4854.52282\tablenotemark{$\dagger$}   &  165.48  &  130.05 \\
 4424.63882  &  180.62  &  114.55 &  4854.52282  &  165.49  &  129.52 \\
 4424.63882\tablenotemark{$\dagger$}   &  180.76  &  114.64 &  5087.61078\tablenotemark{$\star$}  &  145.58  &  149.79 \\
 4470.53847\tablenotemark{$\dagger$}   &  175.79  &  120.98 &  5589.53163  &  172.55  &  123.44 \\
 4470.53847  &  176.07  &  119.67 &  5589.53168\tablenotemark{$\dagger$}   &  172.78  &  123.16  
\enddata
\tablenotetext{$\star$}{HARPS spectra}
\tablenotetext{$\dagger$}{MIKE red spectra}
\end{deluxetable}

\section{Model of the System}
To derive absolute fundamental parameters for the system we used two codes for the analysis of the eclipsing binary stars: the 
Wilson-Devinney (WD) program,
 version 2007 \citep{wil71,wil79,wil90,van07} equipped with the automated Differential Correction (DC) optimizing subroutine; 
 and the JKTEBOP program \citep{sou04a,sou04b} based on the EBOP code \citep{pop81}. The WD code allows to simultaneously solve for
 radial velocity and multi-band light curves recommended as the best way to obtain a consistent model of a binary system, 
 e.g.,~\cite{wil79}, while use of JKTEBOP allows for a reliable determination of the errors by Monte Carlo simulations. The WD code is based 
 on Roche-lobe geometry and employs quite a sophisticated treatment of stellar surface physics, while in the JKTEBOP code stellar surfaces 
 are approximated by biaxial ellipsoids, and simpler description of stellar physics is used. As our system is well detached 
 it is meaningful to compare the results from both programs.

\subsection{Parameter  Choice}
The choice of adjustable parameters has crucial impact on the final solution and parameters of the model. We tried to estimate and fix 
as many parameters as possible. The preliminary orbital period was calculated with the string-length method giving $P=371.71$ days. 
The moment of the primary minimum was estimated to be T$_0=2450776.684$ from the I-band light curve and fixed during the later analysis. 
In this paper we will refer to the primary as the star which is eclipsed in the deeper, primary minimum. The average temperature 
of the primary was estimated as follows. We determined the reddening in the direction of our target to be E(\bv)$=0.06 \pm 0.03$ using 
the OGLE-II reddening maps \citep{uda99} and OGLE-III reddening maps \citep{has11}. Average out-of-eclipse magnitudes were calculated from 
all observations taken outside minima: $V$(15\fm807), $I$(14\fm722), $J$(14\fm020), $K$(13\fm317). These magnitudes were dereddened 
using the interstellar extinction law given by \cite{sch98} and assuming R$_V=3.1$. From combined preliminary solutions obtained 
from RaVeSpAn software written by B.~Pilecki and from JKTEBOP code we obtained the surface gravities: $\log g_1=1.7$ and $\log g_2=1.6$, 
the eccentricity $e=0.31$ and $T_2 - T_1\sim$ 10 K. The metallicity was assumed to be [Fe/H]$=-0.5$ appropriate for a young (age $\sim$ 200 Myr) population of stars in the SMC: e.g. \citet{dia10} from integral spectroscopy of 
 clusters for a cluster NGC 458 (it lies very close to our star)
 derived [Fe/H]$=-0.4$ and age about 100-200 Myr (similar to age of our star). To derive an average effective temperature of the binary we employed a number of empirical calibrations published during last thirteen years 
(Table~\ref{tbl-2}). The resulting average effective temperature was set as a temperature of the primary T$_1=4800 \pm 100$ K. It corresponds to a spectral type G8 giant star according to the calibration by \cite{alo99}. The lower metallicity e.g., [Fe/H]$=-0.7$ has only minor effect on the temperature (it is lower by 10 K).

\begin{deluxetable}{lcc}
\tabletypesize{\scriptsize}
\tablecaption{Effective Temperatures Derived from the Dereddened Colors \label{tbl-2}}
\tablewidth{0pt}
\tablehead{
\colhead{T$_{eff} (K)$} & \colhead{Calibration} & \colhead{Color Index} 
}
\startdata
4732 & \cite{ben98}  & \mbox{$V\!-\!K$} \\
4763 & \cite{alo99}  & \mbox{$V\!-\!K$} \\
4799 & \cite{hou00} & \mbox{$V\!-\!K$} \\
4717 & \cite{ram05} & \mbox{$V\!-\!I$},\mbox{$V\!-\!K$}\\
4830 & \cite{mas06} & \mbox{$V\!-\!K$} \\
4847 & \cite{gon09} & \mbox{$V\!-\!K$} \\
4899 & \cite{cas10} & \mbox{$V\!-\!I$},\mbox{$V\!-\!K$} \\
4812 & \cite{wor11} & \mbox{$J\!-\!K$},\mbox{$V\!-\!I$},\mbox{$V\!-\!K$} \\ 
{\bf 4803} & & {\bf mean}
\enddata
\end{deluxetable}

Using BF we estimated the velocity broadening to be 8.4 km s$^{-1}$ and 10.8 km s$^{-1}$ for the primary and the secondary, respectively, 
with errors $\pm 1.0$ km s$^{-1}$. These velocities are {\it lower} than the appropriate semi-synchronous velocities expected if the components 
are in tidal locking during periastron passage, 12.5 km s$^{-1}$ and 13.2 km s$^{-1}$, respectively. If we account for the macroturbulence 
velocity field typical for a G type giant atmosphere described by $\zeta_{RT}=5.4$ km s$^{-1}$ \citep[Table B.2]{gra05}, the primary velocity 
broadening is consistent with synchronous rotation: $v \sin{i}=6.4$ km s$^{-1}$. However, the secondary seems to rotate super-synchronously with the rotational velocity $v \sin{i}$
being by a factor of 1.4 larger than the synchronous one. Accordingly we set  the star rotation parameters $F_1=1.0$ and $F_2=1.4$. 

The albedo parameter was set to 0.5 and the gravity brightening to 0.32, both values appropriate for a cool, convective atmosphere. 
The limb darkening coefficients were calculated internally by the WD code according to the logarithmic law of \cite{kli70} during each iteration 
of DC using tabulated data computed by \cite{van93}. With the help of BF we tried to detect any signature of a tertiary component in the spectra 
at different orbital phases but we failed, and thus set the third light parameter to $l_3=0$.        

As free parameters of the WD model we chose the semimajor axis $a$, the orbital eccentricity $e$, the argument of periastron $\omega$,  
the phase shift of the primary spectroscopic conjunction $\phi$, the systemic radial velocity $\gamma$, the orbital inclination $i$ , 
the secondary star average surface temperature $T_2$, the modified surface potential of both components $\Omega_1$, $\Omega_2$, 
the mass ratio $q=M2/M1$, the observed orbital period $P_{obs}$, and the relative monochromatic luminosity of the primary star in the two bands $L1_V$, $L1_I$. 
It is worth noticing that within the WD code there is no possibility to directly adjust the radial velocity semiamplitudes $K_1$ and $K_2$, 
instead the semimajor axis and the mass ratio are adjusted simultaneously.    

\subsection{Fitting Procedure}  
We fitted simultaneously two light curves, in the I-band and V-band, and two radial velocity curves using the DC subroutine of the WD code. 
The detached configuration (Mode 2) was chosen during all the analyses and a simple reflection treatment (MREF=1, NREF=1) was employed. 
A stellar atmosphere formulation was selected to both stars (IFAT=1). A level dependent weighting was applied (NOISE=1) and curve dependent 
weightings (SIGMA) were calculated after each iteration. The initial input solution was found using the JKTEBOP code. Convergence was defined 
to have been achieved if the parameter corrections given by DC were smaller than their standard errors on three consecutive iterations. 
The grid size was initially set to N=40 for both stars but other grids were checked. It was found that very fine grids (N$>50$) produced 
some numerical instability and no convergence, in the sense defined above, could be obtained. We did not break the adjustable parameter 
set up into subsets, but instead we adjusted all free parameters at each iteration. 

At the end of the fitting procedure we additionally adjusted the third light $I_3$ to find its impact on the solution: the third light 
corrections were invariably negative suggesting an unphysical solution, and we therefore set $I_3=0$ in our final solution. 

\begin{deluxetable}{lcc}
\tabletypesize{\scriptsize}
\tablecaption{Photometric Parameters \label{tbl-3}}
\tablewidth{0pt}
\tablehead{
\colhead{Parameter} & \colhead{WD\tablenotemark{a}} & \colhead{JKTEBOP\tablenotemark{b}} 
}
\startdata
 Eccentricity $e$ &   $0.3112 \pm 0.0012$ & $0.3064 \pm 0.0010 $ \\
 Argument of Per. $\omega$ (deg)& $30.2 \pm 0.4$  & $28.5 \pm 0.5$ \\
 Phase Shift $\phi$ & $0.07557 \pm 0.00009$ & not adjusted\\
 Orbital Inclination $i$ (deg) & $88.19 \pm 0.02$ & $88.18 \pm 0.02$ \\
 Sec. Temperature $T_2$ (K) & $4813 \pm 3$ & not adjusted \\ 
 Fractional Radius $r_1$ & $0.1097 \pm 0.0007$ & $0.1145 \pm 0.0016$ \\
 Fractional Radius $r_2$ & $ 0.1160 \pm 0.0007$ & $0.1117 \pm 0.0016 $\\
 $(r1+r2)$ &$ 0.2257$& $0.2262 \pm 0.0004 $\\
 $k=r2/r1$& $1.057$& $ 0.975 \pm 0.028 $\\
 Observed Period $P_{obs}$ (d) & $371.768 \pm 0.002$ & $371.768 \pm 0.002$ \\
 $(L2/L1)_V$ & $1.135 \pm 0.011$ & not adjusted \\
 $(L2/L1)_I$ & $1.131 \pm  0.012$ &$ 0.976 \pm 0.052$ \\
 $(L2/L1)_J$ & $1.125$\tablenotemark{c} &$-$ \\
 $(L2/L1)_K$ & $1.120$\tablenotemark{c} &$-$ \\
 Third Light  $I_3$ & $0.0$ & not adjusted
 \enddata
\tablenotetext{a}{Simultaneous solution of V-band and I-band light curves, and radial velocity curves of both componets. Uncertainties quoted are the standard errors from DC subroutine.}
\tablenotetext{b}{Only I-band light curve solution. Uncertainties quoted are from Monte Carlo simulations.}
\tablenotetext{c}{Extrapolated from the model.}
\end{deluxetable}

\begin{deluxetable}{lc}
\tabletypesize{\scriptsize}
\tablecaption{Orbital Parameters \label{tbl-4}}
\tablehead{
\colhead{Parameter} & \colhead{Value} 
}
\startdata
 Periastron Passage (HJD) & $2451114.80$ \\ 
 Orbital Period $P$ (d) & $371.585 \pm 0.002$\tablenotemark{a}\\
 Semimajor Axis $a$ ($R_\sun$) &   $417.49 \pm 0.79$  \\
 Systemic Velocity $\gamma$ (km s$^{-1}$) & $147.68 \pm 0.05$   \\
 Prim. Velocity Semiamplitude $K_1$ (km s$^{-1}$) & $29.93 \pm 0.10$ \\  
 Sec. Velocity Semiamplitude $K_2$ (km s$^{-1}$) & $29.46 \pm 0.11$ \\  
 Mass Ratio $q$ & $1.016 \pm 0.005$ 
 \enddata
 \tablenotetext{a}{Corrected for the radial movement of the centre of the system mass in respect to Solar System Barycenter - see Section~\ref{3.3}.}
\end{deluxetable}

\begin{figure}
\includegraphics[angle=0,scale=.50]{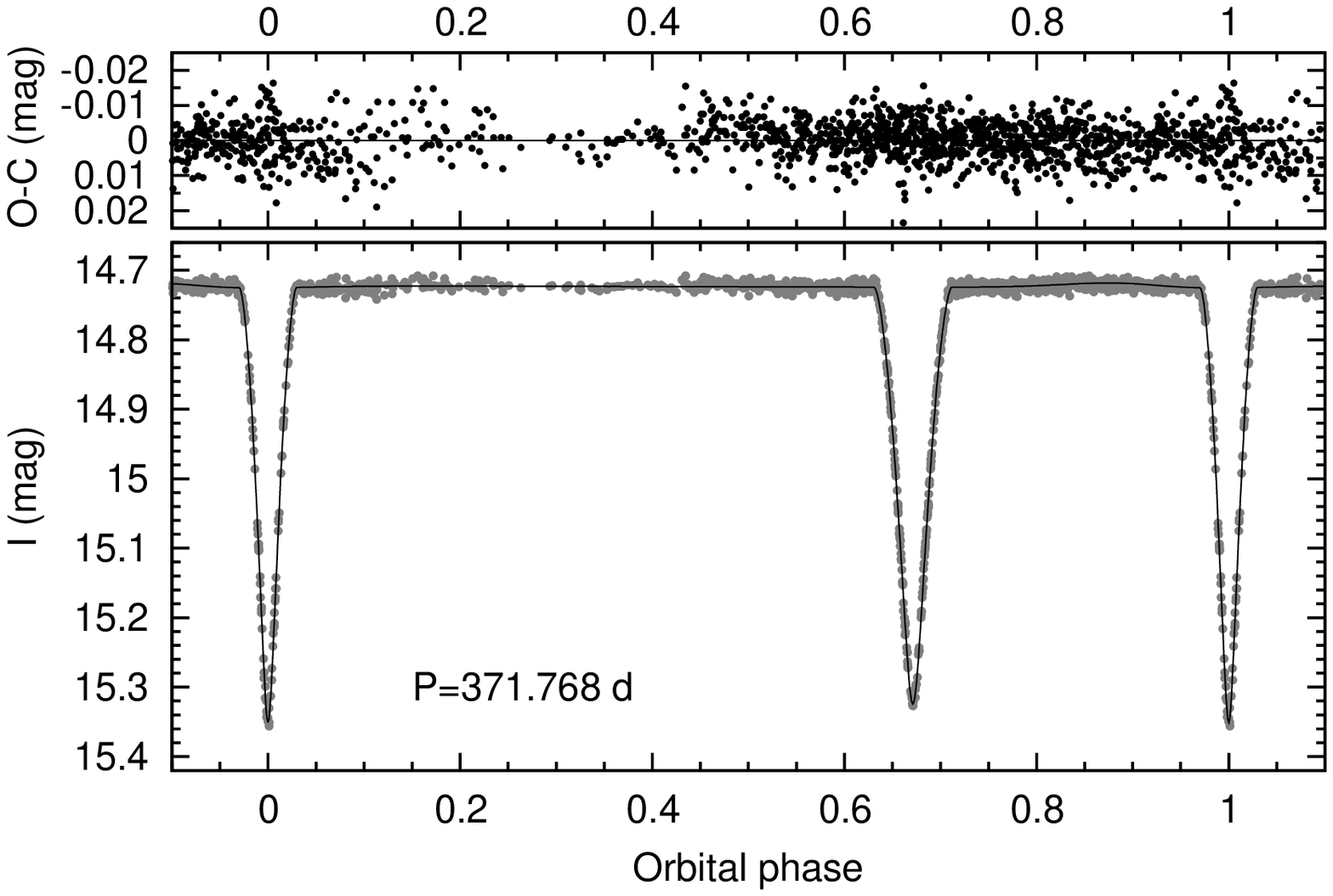}
\caption{ The I-band light curve solution to OGLE SC10 137844 from the WD code.\label{fig1}}
\end{figure}

\begin{figure}
\includegraphics[angle=0,scale=.50]{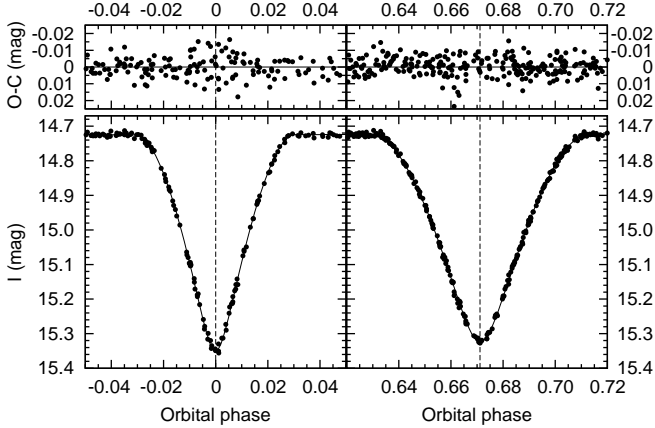}
\caption{ Zoom into the I-band light curve solution to OGLE SC10 137844. Left panel presents the primary minimum, the right panel shows the secondary minimum. \label{fig2}}
\end{figure}

\begin{figure}
\includegraphics[angle=0,scale=.50]{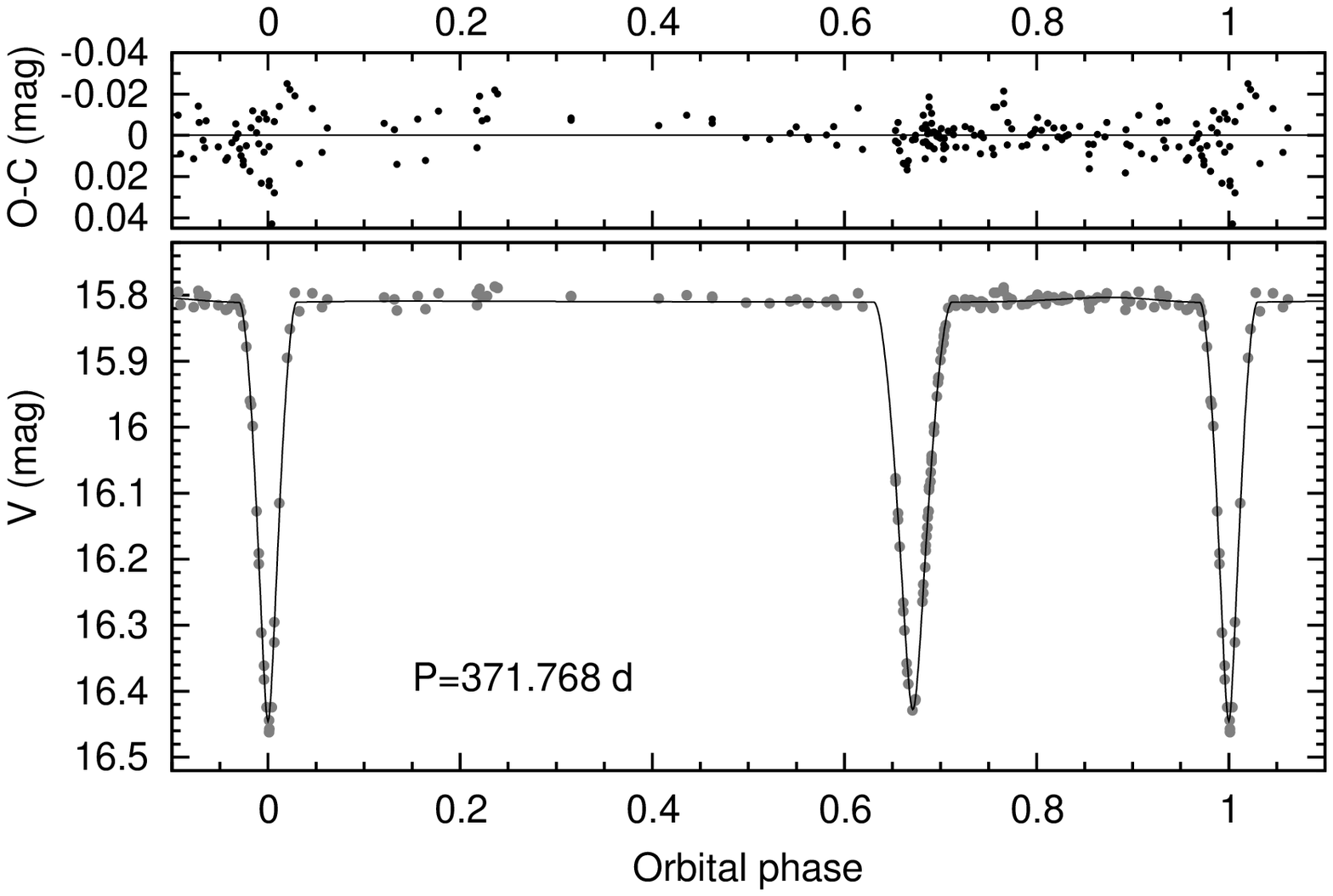}
\caption{ The V-band light curve solution to OGLE SC10 137844 from the WD code.\label{fig3}}
\end{figure}

\begin{figure}
\includegraphics[angle=0,scale=.50]{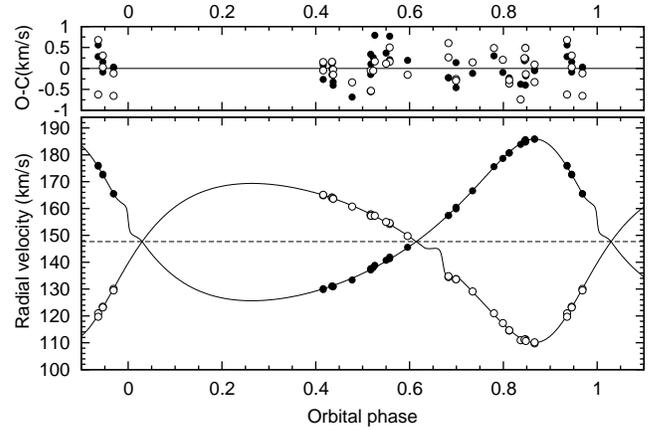}
\caption{ The radial velocity curve solution to OGLE SC10 137844 from the WD code. Filled black circles denote the primary's velocities. The dashed line marks the systemic velocity. The deviation of the secondary radial velocities from the Keplerian motion near the orbital phase 0.69 indicates the Rossiter-McLaughlin effect.\label{fig4}}
\end{figure}

The I-band light curve solution obtained with the WD code is presented in Figures~\ref{fig1} and~\ref{fig2}. This solution was compared 
with the solution resulting from the application of the JKTEBOP code. Using JKTEBOP we fitted only the I-band light curve, and values of the mass ratio and the 
limb darkening coefficients were set according to the final WD solution. The errors of the parameters were calculated from ten thousand 
Monte Carlo simulations (Mode 8 of JKTEBOP code). It turned out that the parameters obtained from both solutions are consistent with each other 
within $2\sigma$ errors  with one exception: the individual fractional radii of the components. The WD solution prefers the secondary star being 
the larger and more luminous of the two stars, while the JKTEBOP solution prefers the opposite. It underlines the old problem of determining 
univocally stellar radii for eclipsing binaries having partial eclipses. To distinguish between the solutions we investigated integrated CCF 
and BF power profiles of both stars to derive their spectroscopic light ratio. The calculated secondary to primary light ratio from CCF and BF 
turned out to be about 1.1. Thus, for consistency of the model, we decided to adopt the radii delivered by the WD code but with uncertainties as 
returned from the JKTEBOP code. The resulting ephemeris for the primary minimum is:
\begin{eqnarray}
T_0 ({\rm HJD}) & = & 2450776.68 (\pm 0.01) + E\!\times\!371.7684 (\pm 0.0017) \nonumber
\end{eqnarray}

The V-band solution is shown in Figure~\ref{fig3}, and the radial velocity curve solution in Figure~\ref{fig4}. The photometric 
and orbital parameters of the final solution are given in Table~\ref{tbl-3} and Table~\ref{tbl-4}, respectively. We note 
that the systemic velocity of our system is in excellent agreement with the mean radial velocity of the red giant branch stars derived 
by \cite{har06} from their IMACS field 3. As that field encompasses the eclipsing binary, the systemic velocity independently confirms 
that our target belongs to the SMC.        

Two of our spectra were taken during secondary eclipse egress and have allowed an approximate modeling of the Rossiter-McLaughlin effect.
 The observed moderate profile (see Figure~\ref{fig4}) was fitted by adjusting the secondary star rotation parameter and 
 we obtained $F_2=1.3 \pm 0.1$. This value seems to confirm that the secondary rotates super-synchronously and it is in accordance 
 with the initial value $F_2=1.4$ derived from the BF analysis.

\subsection{Absolute Dimensions}\label{3.3}
The observed orbital period of the system $P_{obs}$ and the true orbital period $P$ are linked through a relation:
\begin{equation}
P_{obs}  = P (1 + \frac{\gamma}{c}),
\end{equation}
where $c$ is the velocity of light. The corrected period $P$ is listed in Table~\ref{tbl-4} and is used to calculate semimajor axis of the system. Although the correction is small it is much larger than our precision of period determination. 
 
\begin{figure}
\includegraphics[angle=0,scale=.50]{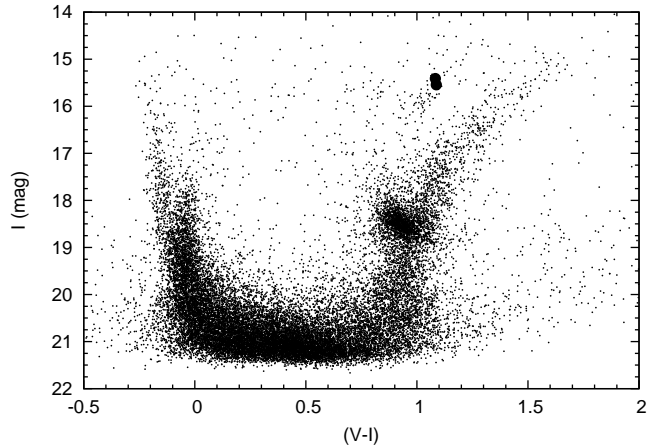}
\caption{ The observed color-magnitude diagram of the OGLE field SMC113.3 containing the eclipsing binary SC 10 137844. Superimposed (full circles)
 are the positions of its two giant components. Both components of our system are clump giants 
located at the red loop of the helium burning. 
At this quiet and relatively long lasting evolutionary 
stage stars do not show significant 
photometric and spectroscopic variations, which make them very good 
targets for precise determination of their stellar parameters.\label{fig5}}
\end{figure}

Once the semimajor axis $a$ of the system is known it is possible to obtain the absolute dimensions of the stars.  Table~\ref{tbl-5} gives astrophysical data about the two components. The observed individual magnitudes and colors were calculated from 
the out-of-eclipse mean magnitudes and the light ratios given in Table~\ref{tbl-3}. The position of the disentangled components on the 
color-magnitude diagram (CMD) is shown in Figure~\ref{fig5}. The physical radii of the stars result from the relation: $R=r\cdot a$, 
where $r$ is the fractional radius listed in Table~\ref{tbl-3}. The masses are derived from the equations:

\begin{eqnarray}
M_1 [M_\sun] & = & 1.34068\cdot 10^{-2}\frac{1}{1+q} \frac{a^3[R_\sun]}{P^2[{\rm d}]}\\
M_2 [M_\sun] & = & M_1\cdot q
 \end{eqnarray}

\begin{deluxetable}{lcc}
\tabletypesize{\scriptsize}
\tablecaption{Physical Properties of the OGLE SC10 137844 System \label{tbl-5}}
\tablewidth{0pt}
\tablehead{
\colhead{Property\tablenotemark{a,b}} & \colhead{The Primary} & \colhead{The Secondary} 
}
\startdata
 Spectral Type  &  G8 II-III & G8 II-III  \\
 $V$ (mag) & 16.630  & 16.493  \\
 \mbox{$V\!-\!I$} (mag) & 1.087 & 1.083 \\
 \mbox{$V\!-\!K$} (mag) & 2.498 & 2.483 \\
 \mbox{$J\!-\!K$} (mag) & 0.705 & 0.701 \\
 Radius ($R_\sun$) & $45.8 \pm 0.7$& $48.4 \pm 0.7$\\
 Mass ($M_\sun$) & $3.504\pm 0.028$ & $3.561 \pm 0.025$ \\
 $\log g$ (cgs) & $1.660 \pm 0.017$ & $1.620 \pm 0.016$ \\
 $T_{\rm eff}$ (K)& $4800 \pm 100$ & $4813 \pm 100$ \\
 $v \sin i$ (km s$^{-1}$) & $6.4 \pm 1.0$&  $8.7 \pm 1.2$\\
 Luminosity ($L_\sun$) & $1000 \pm 80$& $1130 \pm  80$ \\
 $M_{bol}$ (mag) &$-2.75$ & $-2.88$ \\
 $M_V$ (mag) &$-2.40$ & $-2.54$   \\
 Fe/H & \multicolumn{2}{c}{ $-0.5$ dex (assumed)} \\
 E(\bv) & \multicolumn{2}{c}{$0.06 \pm 0.03$}
 \enddata
 \tablenotetext{a}{Absolute dimensions were calculated assuming: $G=6.673\cdot10^{-8}$ cm$^3$g$^{-1}$s$^{-2}$, $R_\sun=6.9551\cdot10^{10}$ cm, $M_\sun=1.9888\cdot10^{33}$ g, $M_{bol,\sun}=+4.75$.}
\tablenotetext{b}{The magnitudes and colors are observed values.}
\end{deluxetable}

 Derived absolute visual luminosities of both componets are consistent with luminosity class II-III for G-type star \citep{egg94} and their masses lie in the mass range expected for bright giant stars which have masses from about 3 $M_\odot$ to 9 $M_\odot$.

\section{Distance to the System}
To derive the distance of our binary system we employed a surface brightness - color relation.  We used the calibration of \cite{ben05} between 
$V$-band surface brightness and \mbox{$V\!-\!K$} color relation (SBR) obtained from a number of carefully selected mixed sample 
of giant and dwarf stars having their angular diameters measured precisely by long baseline interferometry. The angular diameter of a star 
predicted by the surface brightness-color relation takes the form:
\begin{eqnarray}
\phi\;[{\rm mas}] & = & 10^{0.2\cdot({S-m_0})},
 \end{eqnarray}
where $S$ is the surface brightness in a given band and $m_0$ is the unreddened magnitude of a given star in this band. The distance in parsecs follows then directly from angular diameter scaling and is given by a simple linear equation:
 \begin{eqnarray}
d\;[{\rm pc}] & = & 9.2984 \cdot \frac{R\;[R_\sun]}{\phi\;[{\rm mas}]}\label{eqn4} 
 \end{eqnarray}
The distance to the OGLE SC10 137844 system was calculated as the average distance to both components, and its value is $58.4 \pm 1.8$ kpc 
corresponding to a true distance modulus of $18.83 \pm 0.06$ mag. The distance moduli of the primary and the secondary differ by just 0.001 mag, 
serving as an independent check on the model consistency.  We compare our result with distance estimates obtained from other recent 
surface brightness - color calibrations, results being summarized in Table~\ref{tbl-6}.  Within the errors all distances are consistent with our 
adopted value based on the calibration by \cite{ben05}. 

\begin{deluxetable}{lccc}
\tabletypesize{\scriptsize}
\tablecaption{Distance determinations to OGLE SC10 137844 from different surface brightness-color relations\label{tbl-6}}
\tablewidth{0pt}
\tablehead{
\colhead{$d$ [kpc]} & \colhead{Luminosity Class} & \colhead{Color} & \colhead{Reference} 
}
\startdata
$57.5 \pm 2.0$ & Dwarfs+Subgiants  & \mbox{$V\!-\!K$} & \cite{ker04} \\
$58.0 \pm 2.8$ & Giants & \mbox{$V\!-\!K$} & \cite{gro04} \\
$61.6 \pm 4.2$ & Dwarfs & \mbox{$V\!-\!I$} & \cite{ker08} \\
$60.8 \pm 2.0$ & Dwarfs+Subgiants & \mbox{$V\!-\!K$} & \cite{cas10} \smallskip \\
{\bf 58.4}$\pm${\bf 1.8} & {\bf Giants} & {\bf \mbox{$V\!-\!K$}} & {\bf \cite{ben05}}
\enddata
\end{deluxetable}

\subsection{Error Budget}
The systematic uncertainty on the surface brightness relation itself provides the largest contribution to the total error of 2.0\%, 
(compare with e.g.,~\cite{pie09}). The metallicity dependence of the SBR is very weak, for our system we expect a correction of about 
0.006 mag \citep{tom01,ben98}, and we decided to add an additional error of 0.3\% in the error budget.  Systematic zero point uncertainties of absolute photometry are 0.7\% for near infrared SOFI bands and 0.5\% for the optical OGLE $V$ band \citep{uda00}.  The estimated total systematic error is 2.4\%.

The absolute radii are known with an accuracy of 1.5\%. It would mean that the contribution to the distance error from the radius uncertainty 
was 1.5\%. However, because of the similar diameters and temperatures of the components this contribution is much smaller and, to the 
first order of magnitude, equal to the uncertainty of the sum of the radii. Following Equation~\ref{eqn4} the average distance to the system 
can be written as:
\begin{eqnarray}
\!\!\!\!\!\!\! \langle{\rm d}\rangle  =  \frac{k}{2} \left( \frac{R_1\phi_2 + R_2\phi_1}{\phi_1\phi_2} \right) & = & \frac{k}{2} \left( \frac{R_1+R_2}{\phi_2} + \frac{R_1}{\phi_2}\frac{\Delta\phi}{\phi_1}\right) 
\label{eqn5}
\end{eqnarray} 
where index 1,2 denotes, the primary and the secondary star, respectively, $\Delta\phi=\phi_2-\phi_1$ and $k=9.2984$. The second term 
in Equation~\ref{eqn5} constitutes, in our case, only about 3\% of the first term. Regarding the error contribution from the radius
uncertainties the first term gives  $\sigma_{(R_1+R_2)}/(R_1+R_2) = 0.35/94.3$ i.e.~about 0.4\% and the second term gives 0.05\%. Combining it with the semimajor axis uncertainty of 0.2\% results in a total uncertainty of 0.5\%.  

The accuracy of $V\!-\!K$ color determination is about 0.015 mag adding 0.7\% to the statistical error.
The disentangling of individual V-magnitudes and near infrared color indexes contributes only slightly to the total error due to the 
similar temperatures of the stars. We recalculated the distance modulus of the system by enlarging the light ratios in the $V$ and $K$ bands 
with the error quoted in Table~\ref{tbl-3} and an error of 0.014, respectively. The resulting distance moduli differ by just 0.0005 mag. 
Thus we omitted this uncertainty in the error budget. 

In the fitting procedure we set atmosphere model formulation in the WD code. We have checked how the distance estimate will change if we assume 
a blackbody approximation. We re-fitted the data setting IFAT=0 (black body) to both stars. The resulting radii and light ratios lead to a value
of the distance modulus which differs by less than 0.001 mag from our adopted value.

The interstellar absorption, 
because of the reddening vector being almost parallel to the SBR, contributes relatively little to the total error: 0.7\%. 
A different interstellar extinction law with~$R_V=2.7$ contributes insignificantly on the level of 0.3\%. Adding the contributions 
of these errors on the distance determination to OGLE SC10 137844 quadratically we obtain a total statistical error of 1.1\%. 

 We conclude that the uncertainty of the extinction and $V\!-\!K$ color determination are the main contributors to the statistical error in the distance. The calculation 
 of the total uncertainty is a bit ambiguous 
 because it depends on the systematic uncertainty probability distribution which we don't know {\it a priori}. We can give however 
 the limits for the total error. The lower error limit of 2.5\% comes from the assumption of a gaussian distribution 
 of the systematic uncertainty while the upper limit of 3.5\% results from the assumption of a pure bias. The mean value of both 
 was given in Table~\ref{tbl-6} corresponding to the total uncertainty of $\sim\!3.0$\%.

\section{Discussion}

We obtain a true distance modulus for our binary system of 18.83 $\pm$ 0.02 (statistical) $\pm$ 0.05 (systematic from the SBR) mag. 
Our analysis demonstrates that for such late-type systems consisting of red giant stars we can indeed measure the distances 
very accurately, with prospects of improving the achievable accuracy once an improved surface brightness-color relation becomes available which is the current limiting factor on the precision of the method. 

In Table 7 we have compiled other recent determinations of the distance to the SMC. Our result from SC10 137844 is in good agreement
with the average SMC distances derived from classical and Type-II Cepheid, and RR Lyrae star samples. The outlier is the distance
of 19.11 mag reported by \cite{nor10} from an analysis of 33 eclipsing binaries in the southwest part of the galaxy. However, 
since the SMC is a line-of-sight elongated structure, especially towards the northeast, with a typical depth 
of $\sim\!0.25$ mag (e.g.~\citet{har06}), all the determinations in Table 7 might be compatible within their quoted errors. 
Also it is known that there is a distance gradient over the SMC with its northeastern part being closer to us 
than its southwestern part (e.g.~\cite{sub11}). Indeed, the distance to the galaxy centre given by \cite{hil05} lies between our determination 
and the \cite{nor10} estimate which seems to be consistent with the present geometrical model of the galaxy. However, because \cite{nor10} did not give 
any estimate of the systematic uncertainty of their result it is not easy to compare the SMC distances obtained from eclipsing binaries.   
 
\begin{deluxetable}{lcccc}
\tabletypesize{\scriptsize}
\tablecaption{Comparison of recent distance determinations to the SMC.\label{tbl-7}}
\tablewidth{0pt}
\tablehead{
\colhead{$(m-M)$} & \colhead{Stat. Err.} & \colhead{Sys. Err.} & \colhead{Method}  & \colhead{Ref\tablenotemark{a}} 
}
\startdata
18.91 & 0.04 & $\sim\!0.1$ & Eclipsing Binaries & 1 \\ 
18.96 & 0.10 & ? & RR Lyr & 2  \\
18.93 & 0.02 & ? & Bump Cepheids & 3 \\
18.97 & 0.03 & 0.12 & RR Lyr & 4  \\
18.85  & 0.07& $\sim\!0.1$& Type-II Cepheids  & 5  \\
19.11 & 0.03 & ? & Eclipsing Binaries &  6  \\
18.92 & 0.14 & ? & Classical Cepheids  & 7 \\
18.83 & 0.02 & 0.05 & Eclipsing Binary  & 8  
\enddata
\tablenotetext{a}{References: 1-  \cite{hil05},  2-  \cite{st06}, 3 -  \cite{kel06}, 4 - \cite{sze09}, 5 - \cite{cie10}, 6 - \cite{nor10}, 7 - \cite{sto11b}, 8  - this paper}
\end{deluxetable}
If we assume a distance gradient over the SMC projected disc of 2-3 kpc \citep{hil05} our star should be 
around $\sim\!1$~kpc (0.035 mag) closer to us than the optical center of the galaxy. Thus our distance would correspond 
to a distance modulus of 18.87 mag for the SMC center. Assuming a modulus offset between SMC and LMC equal to 0.44 \citep{cio00} it would 
imply a distance modulus of the LMC of 18.43 mag. This value is very close to the true LMC distance modulus derived recently 
by \cite{sto11b} with the near-infrared Baade-Wesselink method applied to classical Cepheids.

We are currently working on a number of similar late type eclipsing systems in the Small Magellanic
Cloud, and expect to determine the mean distance to the SMC with a precision better than 4\%.

\acknowledgments
DG, GP, WG, BP, RM and DM gratefully acknowledge financial support for this work from the Chilean Center for Astrophysics FONDAP 15010003, and from the BASAL Centro de Astrofisica y Tecnologias Afines (CATA) PFB-06/2007. Support from the Polish grant N203 387337, Ideas Plus program of 
Polish Ministry of Science and Higher Education, and the 
FOCUS and TEAM subsidies of the Foundation for Polish Science (FNP) 
is also acknowledged. We greatly appreciate the expert support
of the ESO staff at the ESO La Silla Observatory and of the Las Campanas Observatory staff. We thank also the anonymous referee for comments which helped to make this paper more clear. 

This work was supported by the National Science Foundation under grant 
AST-1008798 to RPK and FB. Moreover, RPK acknowledges support by the 
Alexander-von-Humboldt Foundation and the hospitality 
of the Max-Planck-Institute for Astrophysics in Garching and the 
University Observatory Munich, where part of this work was carried out.

{}    

\end{document}